\documentclass[aps,amsmath,showpacs,amsfonts,10pt]{revtex4}
\usepackage{epsfig,graphicx}
\usepackage[english]{babel}
\usepackage{amsfonts}
\usepackage{amsmath}
\usepackage{latexsym}
\usepackage{graphics,bm}
\usepackage{natbib}
\usepackage{dcolumn}
\usepackage{bm}
\usepackage{rotating}
\usepackage{epstopdf}
\begin{document}

\title{Influence of periodically modulated cavity field on the generation of atomic-squeezed states}

\author{Neha Aggarwal$^{1,2}$, Aranya B Bhattacherjee$^{3}$, Arup Banerjee$^{4}$ and Man Mohan$^{1}$}

\address{$^{1}$Department of Physics and Astrophysics, University of Delhi, Delhi-110007, India} \address{$^{2}$Department of Physics, ARSD College, University of Delhi (South Campus), New Delhi-110021, India}\address{$^{3}$School of Physical Sciences, Jawaharlal Nehru University, New Delhi-110067, India}\address{$^{4}$BARC Training School at RRCAT and Homi Bhabha National Institute, Raja Ramanna Centre for Advanced Technology, Indore 452013, India}

\begin{abstract}
We investigate the influence of periodically time-modulated cavity frequency on the generation of atomic squeezed states for a collection of N two-level atoms confined in a non-stationary cavity with a moving mirror. We show that the two-photon character of the field generated from the vacuum state of field plays a significant role in producing the atomic or spin squeezed states. We further show that the maximum amount of persistent atomic squeezing is obtained for the initial cavity field prepared in the vacuum state.
\end{abstract}


\maketitle

\section{Introduction}
It is well known that non-classical state of electromagnetic field like squeezed state can be generated from vacuum in a non-stationary cavity with oscillating wall. This phenomenon of photon generation from vacuum state by non-adiabatic change in the boundary condition of quantum fields is termed as Dynamical Casimir Effect (DCE). Several theoretical studies devoted to the generation of photons in a cavity with vibrating wall resulting into harmonic modulation of the mode frequency have been reported in the literature \cite{cklaw,dodonov1995,dodonov1996,plunien2000,crocee2001}. We also refer to  Refs. \cite{dodonov2010} and \cite{nation2012} for review on various aspects of DCE. It has been shown that for an ideal cavity (without cavity dissipation) maximum number of photons are created when the modulation frequency is twice that of unperturbed cavity mode frequency. Moreover, under this resonance condition the non-classical nature like squeezing property also increases with time. From experimental point of view the major hurdle in generating photons by DCE lies in realizing vibrating cavity with frequency of the order of few GHz. Furthermore, the detection of the generated photon requires coupling the field mode to a detector, which can in turn alter the statistical property of the radiation field significantly due to back action.  In order to investigate the effect of back action on the generated photons, Dodonov and co-workers have carried out series of studies on the interaction of various model atoms with the field in a cavity with harmonically modulated mode frequency \cite{dodonov1995,dodonov2011,dodonov20121,dodonov20122,dodonov20123}. Interaction of a single two-level atom with quantized electromagnetic field inside a cavity with time dependent parameters has also been studied in Ref. \cite{lozovik2001}. Keeping these difficulties in mind recently two proposals for experimental generation and detection of photons in non-stationary cavities have been reported in Refs. \cite{yamamoto} and \cite{onofrio}. In Ref. \cite{onofrio} authors proposed that cavities with mechanical vibration in the GHz range may be obtained through a film bulk acoustic resonator (FBAR) \cite{ruby,cleland} made of vibrating aluminium nitride film of thickness equal to half of the acoustic wavelength. For detection of photons generated in the cavity these authors proposed to use ultra-cold alkali atoms in their hyperfine states. On the other hand, Ref. \cite{yamamoto} explored the possibility of using a cavity with plasma mirror made from semiconductor slab irradiated by periodic laser pulses for generation of photons from vacuum state and Rydberg atoms with large values of principal quantum number for detection of these photons.

One of the hallmarks of the photon generated in the non-stationary cavity is the presence of two-photon correlation as they are generated via two-photon process. It is then natural to explore the possibility of transferring these two-photon correlations to a collection of atoms to generate so-called correlated atomic squeezed (spin squeezed) states \cite{kita,wine,wine1,toth,nori} or atomic entangled states. To this end in this paper we study the interaction of N two-level atoms with a quantized single radiation mode of a non-stationary cavity with harmonically oscillating wall. In particular we focus our attention on the effect of modulation of the cavity mode frequency due to harmonic oscillation of the cavity mirror on the squeezing properties of an ensemble of atoms interacting with this cavity mode. Here we wish to note that dynamics of N-two level atoms in a non-stationary cavity has already been investigated in the context of detection of photons generated in the cavity \cite{dodonov20123,yamamoto}. On the other hand, in the present work we concentrate on the generation of atomic squeezed state in a non-stationary cavity. We should also remark here that several early proposals for the  generation of atomic squeezed states involved transfer of two-photon correlation from radiation mode to the collection of atoms by using photon-atom interaction \cite{nori}. In accordance with this idea the generation of atomic or spin squeezed state by the transfer of two-photon correlation from multi-mode squeezed vacuum to an ensemble of atoms have already been reported in the literature \cite{palma,puri,arup,vernac}.

Before proceeding further we wish to mention that atomic squeezed states are useful quantum resources to improve the precision of measurements in experiments \citep{wine,wine1,polzik,cronin} and to study the particle correlations and entanglement \citep{sorensen,bigelow,gue}. These states have found application in atomic clocks for reducing quantum noise \citep{wine,wine1,tur,meyer,leib} and quantum information processing \citep{sorensen,wang,kor,yi}. The utility of atomic squeezed states in spectroscopy and meteorology has motivated us to explore the possibility of generation of these states in a non-stationary cavity. The rest of the paper is organized in following manner. In section II we give description of the model used in this paper. The results are presented and discussed in section III and the paper is concluded in section IV.
 
\section{Model Hamiltonian}

In order to study the generation of atomic squeezed or spin squeezed states by DCE we consider a collection of $N$ two-level atoms interacting with a quantized single mode of radiation field of a cavity with an oscillating wall. Following Ref. \cite{dodonov20123}, the Hamiltonian for atom plus non-stationary cavity system under rotating wave approximation is written as ( we use $\hbar=1$ throughout the paper) 
\begin{equation}\label{eq1}
H = \omega_{a}J_{z}+\omega_{c}(t)a^{\dagger}a + g_{0}\left (aJ_{+} + a^{\dagger}J_{-}\right ) + i\xi(t)(a^{\dagger 2}-a^{2}).
\end{equation}
where $a$ and $a^{\dagger}$ are the cavity annihilation and creation operators respectively, satisfying the commutation relation $[a,a^{\dagger}] = 1$. The ensemble of $N$ atoms is described using the picture of a collective spin operators $J_{z} = \sum_{i}^{N}(|e_{i}\rangle\langle e_{i}| - |g_{i}\rangle\langle g_{i}|)$, $J_{+} = \sum_{i}^{N}|e_{i}\rangle\langle g_{i}|$, and $J_{-} = \sum_{i}^{N}|g_{i}\rangle\langle e_{i}|$, where $|e_{i}\rangle$ and $|g_{i}\rangle$ represent the excited and the ground states of the \textit{i}th two-level atom, respectively. 
The spin operators are dimensionless and satisfy the commutation relations $[J_{+},J_{-}]=2J_{z}$ and $[J_{\pm},J_{z}]=\mp J_{\pm}$. The Hilbert space of these atomic operators is spanned by symmetric Dicke states $|J,M\rangle$ with $M=-J,-J+1......J-1,J$ \cite{arecchi}. The total spin length is given by $J=N/2$. The Dicke states are eigenstates of $J^{2}$ and $J_{z}$ such that $J_{z}|J,M\rangle = M|J,M\rangle$ and $J^{2}|J,M\rangle = J(J+1)|J,M\rangle$. The lowering and raising operators act on these states as $J_{\pm}|J,M\rangle = \sqrt{J(J+1)-M(M\pm1)}|J,M\pm1\rangle$. 
The parameters $\omega_{a}$, $\omega_{c}$, and $g_{0}$ denote the atomic transition frequency, cavity mode frequency, and atom-field coupling constant (which is assumed to be real) respectively. The harmonic time dependence of the cavity mode frequency $\omega_{c}(t)$ and the last term, which is nonlinear in $a$ and $a^{\dagger}$ arise due to harmonic motion of the cavity boundary \cite{cklaw}. A brief derivation of this nonlinear part of the Hamiltonian is presented in Appendix A.  For the purpose of calculations, following earlier works, we assume that the cavity mode frequency has sinusoidal dependence given by $\omega_{c}(t)=\omega_{0}(1+\epsilon \sin(\Omega t))$ with the unperturbed frequency $\omega_{0}$, which is set to $1$ for simplicity and $\epsilon$ and $\Omega$ representing the modulation amplitude and the modulation frequency respectively. 
Furthermore, note that there are basically two kinds of non-adiabatic processes occurring in a non-stationary  cavity system described by the above Hamiltonian \citep{cklaw}. The first kind is characterized by the $a^{\dagger}a$ term in which the total number of photons inside the cavity does not change. Such a process is known as the zero-photon process. On the other hand, the second kind is represented by the terms $a^{\dagger 2}$ and $a^{2}$ which are responsible for the generation of squeezed photons from the vacuum state. The last term in the Hamiltonian introduces two-photon correlation in the cavity mode and in this paper we explore the possibility of transferring this correlation from photons to atoms via atom-photon interaction.  Also, $\xi(t)$ is the effective frequency which is a arbitrary function of time and is related to $\omega_{c}(t)$ as \cite{cklaw}:
\begin{equation}\label{eq2}
\xi(t)=\frac{1}{4\omega_{c}(t)}\frac{d \omega_{c}(t)}{dt}.
\end{equation}
Considering the realistic case of small-amplitude time modulation i.e., $|\epsilon|<<1$, one can obtain the following expression for $\xi(t)$ from eqn.(\ref{eq2}):

\begin{equation}\label{eq3}
\xi(t) = 2\xi_{0}\cos(\Omega t),
\end{equation} 
where $\xi_{0}=\epsilon \Omega/8<<1$. 
Note that for $\omega_{c}(t)$ independent of time the coefficient $\xi(t) = 0$ and the above Hamiltonian reduces to Tavis-Cummings Hamiltonian \cite{tavis} which has been extensively studied in the context of cavity-QED.

Having discussed the basic model of multi-atom system coupled to a field mode of a non-stationary cavity we now briefly describe the squeezing parameter employed in this paper to characterize the atomic squeezing. We note here that several definitions for the spin or atomic squeezing parameters have been proposed in the literature in different contexts \cite{kita,wine,wine1,toth,nori}.  For example, according to the definition of Kitagawa and Ueda, a state is spin squeezed only if the variance of one spin component $J_{\perp}$ normal to the mean spin vector is less than the variance for a Bloch state ($J/2$) \cite{kita}. In accordance with the definition of Kitagawa and Ueda,  spin squeezing parameter $\zeta_{S}$ is written as \cite{kita}:

\begin{equation}\label{eq7}
\zeta_{S}=\sqrt{\frac{\min(\Delta J_{\vec{n}_{\perp}}^{2})}{J/2}}=\sqrt{\frac{4 \min(\Delta J_{\vec{n}_{\perp}}^{2})}{N}},
\end{equation} 
where the subscript $\vec{n}_{\perp}$ refers to an axis perpendicular to the mean-spin direction $\vec{n}_{0}=\langle \vec{J}\rangle/|\langle \vec{J}\rangle|$ and the minimization is over all directions $\vec{n}_{\perp}$. This parameter is used to quantify the degree of quantum correlations among the elementary spins. The atomic-squeezing condition in terms of this parameter is given by the condition $\zeta_{S}<1$, i.e., the fluctuation in one direction is reduced. Indeed, it has a very close relation with quantities such as concurrence \cite{wang} and negative correlations \cite{ulam}.
On the other hand, the squeezing parameter proposed by Wineland et al. \cite{wine} was in the context of Ramsey spectroscopy for the determination of transition frequency and consequently this parameter is also termed as spectroscopic squeezing. This squeezing parameter is related to the ratio of fluctuations in the measurement of resonance frequency using an ensemble of atoms in a general atomic state and in a coherent spin state. The spectroscopic squeezing parameter $\zeta_{R}$ is given by
\begin{equation}\label{eq8a}
\zeta_{R}=\sqrt{N\frac{\min(\Delta J_{\vec{n}_{\perp}}^{2})}{|\langle\vec{J}\rangle|^{2}}},
\end{equation} 
These two squeezing parameters are related to each other and it can be seen from their definitions $\zeta_{S}^{2} \leq \zeta_{R}^{2}$.  In this paper we employ both the squeezing parameters mentioned above to characterize the squeezing of property of a collection of atoms interacting with field mode of a non-stationary cavity. To this end we need to calculate averages of angular momentum operators and their second-order moments in the  combined state $|\Psi_{com} (t)\rangle$ of atoms plus field satisfying time-dependent Schrodinger equation 

\begin{equation}
\label{schr}
\frac{d |\Psi_{com} (t)\rangle}{dt} = -i H |\Psi_{com} (t)\rangle.
\end{equation}
In order to solve this differential equation we make use of the ansatz 
\begin{equation}
|\Psi_{com}(t)\rangle = \sum_{n,M} C_{n,M}(t)|n\rangle |M\rangle,
\end{equation}

where $C_{n,M}$ are the time-dependent coefficients and $|n\rangle$ represents the number state, which is an eigenstate of number operator $a^{\dagger}a$ such that $a|n\rangle = \sqrt{n}|n-1\rangle$ and $a^{\dagger}|n\rangle = \sqrt{n+1}|n+1\rangle$. For convenience we use the notation $|M\rangle  = |J,M\rangle$ since $J^{2}$ is constant of motion for $H$ given by Eq. (\ref{eq1}). We further assume that the radiation mode and the collection of atoms are initially uncorrelated and the initial state of the atom plus field system can be written as a direct product
\begin{equation}
|\Psi_{com}(0)\rangle = \left(\sum_{n}c_{n}|n\rangle\right)\otimes|\psi(0)\rangle,
\end{equation}
where $|\psi(0)\rangle = |J=N/2,M=-N/2\rangle$ represents the initial atomic state in which all the atoms are occupying their ground states and the coefficients $c_{n}$ are the projections of initial field state on the number state $|n\rangle$.  Using Schrodinger's equation (Eq. (\ref{schr})) with Hamiltonian H of Eq.(\ref{eq1}) we write the equation of motion for the coefficient $C_{n,M}$ as

\begin{eqnarray}\label{eq8}
i\dot{C}_{n,M}(t) & = & \left[(1+\epsilon \sin(\Omega t))n+\omega_{a}M \right]C_{n,M}(t)\nonumber \\&+& g_{0}\left[\sqrt{n}\sqrt{J(J+1)-M(M+1)}\right]C_{n-1,M+1}(t)\nonumber \\&+& g_{0}\left[ \sqrt{n+1}\sqrt{J(J+1)-M(M-1)}\right]C_{n+1,M-1}(t)\nonumber \\&+&2i\xi_{0}\cos(\Omega t) [\sqrt{n+1}\sqrt{n+2}]C_{n+2,M}(t)-2i\xi_{0}\cos(\Omega t)[\sqrt{n}\sqrt{n-1}]C_{n-2,M}(t).
\end{eqnarray}

We wish to point out that in general the above differential equation cannot be solved analytically due to coupling of coefficient $C_{n,M}$ with infinitely many $C_{n-2,M}$, $C_{n-4,M}$... and $C_{n+2,M}$, $C_{n+4,M}$..... coefficients. Consequently, one needs to solve these coupled differential equations numerically. To implement the numerical method for practical reasons it becomes necessary to truncate the number state basis of cavity mode. To this end we choose adequate number of basis states $|n\rangle$ and ensure the convergence of results by increasing the basis size. In the next section we present and discuss the results obtained by us. 

\section{Results and Discussions}

We begin this section by presenting the results, which have been obtained by numerically solving Eq.(\ref{eq8}) with $N = 20$ ($J = 10$) and considering initially the atomic and the cavity field mode to be in $|\psi(0)\rangle = |J,-J>$ and vacuum state $|0\rangle$  ($c_{n} = 1$ for $n = 0$ and $c_{n} = 0$ for $n \neq 0$) respectively. The results for the evolution of atomic squeezing parameter $\zeta_{S}(t)$ and $\zeta_{R}(t)$ are displayed in Figs. 1(a) and 1(b) respectively for  various values of modulation amplitude $\epsilon$ with the modulation frequency fixed at $\Omega = 2.0$. This choice of value of modulation frequency is guided by the result that maximum number of squeezed photons are created in an ideal empty non-stationary cavity when the modulation frequency is twice that of unperturbed cavity mode frequency. It can be clearly seen that under this condition both the  atomic squeezing parameters become less than $1$ for non-zero values of the modulation amplitude indicating generation of atomic or spin squeezed state. On the other hand, for $\epsilon = 0$, no atomic squeezing is observed as both $\zeta_{S}(t)$ and $\zeta_{R}(t)$ throughout remain unity. This constitutes the main result of this paper. Moreover, we observe that the results presented in Figs. 1(a) and 1(b) satisfy the inequality $\zeta_{S}^{2} \leq \zeta_{R}^{2}$. The last term in the Hamiltonian with $a^{\dagger 2}$ and $a^{2}$  results in the generation of squeezed state of radiation field from the vacuum state via two-photon process  and the two-photon correlation of squeezed state is transferred to the ensemble of atoms via interaction of atoms with this cavity field mode.  Moreover, from  Figs. 1(a) and 1(b), we observe that persistent atomic squeezing, which lasts for longer time duration is produced under resonant condition ($\Omega = 2\omega_{0}$) and the magnitude of squeezing also increases with the increase in modulation amplitude. This increase in atomic squeezing is attributed to the enhancement in squeezing characteristic of the cavity field with higher values of modulation amplitude $\epsilon$. Thus, the modulation amplitude acts as an additional tool for controlling the degree of atomic squeezing generated by photon-atom interaction in a non-stationary cavity. 

\begin{figure}[h]
\hspace{-0.0cm}
\begin{tabular}{cc}
\includegraphics [scale=0.70]{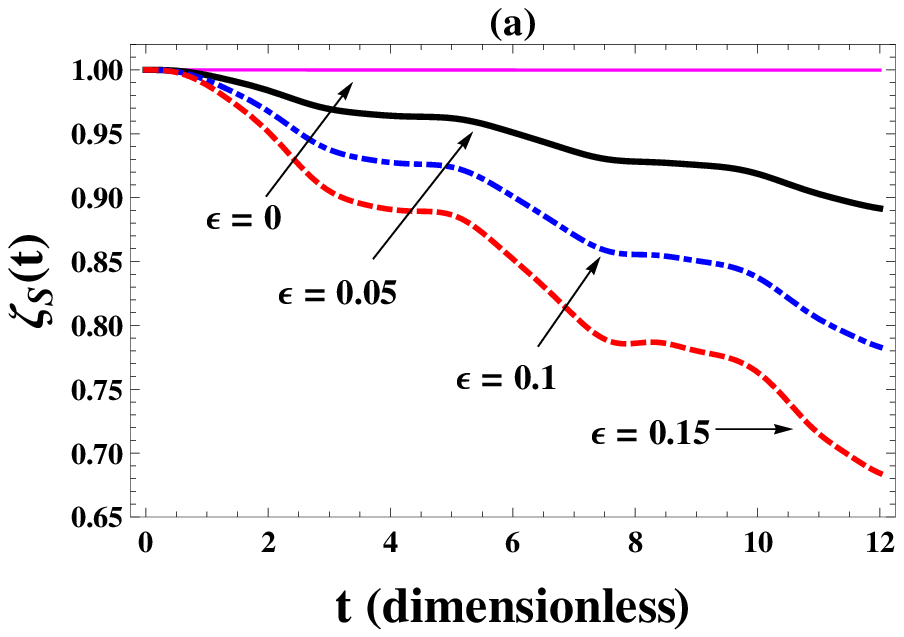}& \includegraphics [scale=0.70]{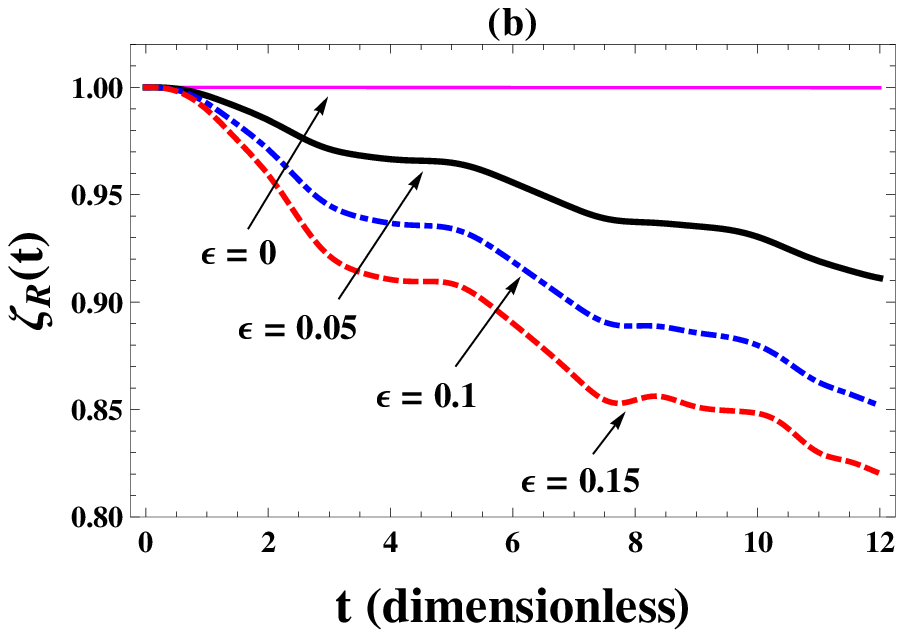}\\
\end{tabular}
\caption{(color online) (a) Plot of squeezing parameter $\zeta_{S}(t)$ as a function of dimensionless time ($t$) for four different values of modulation amplitude with $\epsilon=0$ (thin solid line), $\epsilon=0.05$ (thick solid line), $\epsilon=0.1$ (dot dashed line) and $\epsilon=0.15$ (dashed line). (b) Plot of squeezing parameter $\zeta_{R}(t)$ versus dimensionless time ($t$) for four different values of modulation amplitude with $\epsilon=0$ (thin solid line), $\epsilon=0.05$ (thick solid line), $\epsilon=0.1$ (dot dashed line) and $\epsilon=0.15$ (dashed line). The other parameters used are $g_{0}=0.6$, $\Omega=2$ and $\omega_{a}=1$. We assume $|\psi(0)>=|J,-J>$ ($J=10$) and the cavity field to be initially in the vacuum state.}
\end{figure}\label{fig1}

To get a better insight into the mechanism of the generation of atomic squeezed states via atom-photon interaction in a non-stationary cavity with oscillating wall, we now focus our attention on a system of two two-level atoms interacting with a single quantized cavity mode. The time dependent Schrodinger equation with the model Hamiltonian $H$ given by Eq. (\ref{eq1}) for two atoms ($N = 2)$ can be solved analytically under particular value of detuning between the cavity mode ($\omega_{0}$) and modulation frequency ($\Omega$). Following Refs. \cite{dodonov20122,dodonov20123}, we find that 
for $\Omega = 2 + \delta$ with $\delta = \pm\sqrt{6}g_{0}$ and  both the atoms initially in their ground states ($|g_{1}|g_{2}\rangle$) and the field mode in vacuum state ($|0\rangle$) only four essential atom plus photon states get coupled and acquire significant probabilities of occupation. Under this condition no more than two photons can be generated from the initial state mentioned above. Therefore, for this non-resonant condition the time dependent atom plus photon state for time $t > 0$ can be written as

\begin{equation}
|\Psi_{com}(t)\rangle = a_{0}(t)e^{it}|0,-1\rangle + a_{2}e^{-it}(t)|2,-1\rangle  + b_{1}(t)e^{-it}|1,0\rangle + d_{0}(t)e^{-it}|0,1\rangle
\label{twoatomstate}
\end{equation}
where as mentioned before state $|n, M\rangle$ denote the combined state of photons in number eigenstate $|n\rangle$ and the two-atom system in a collective state $|M\rangle$ with $M = -1, 0, 1$. In terms of the ground ($|g_{1}\rangle$, $|g_{2}\rangle$) and excited ($|e_{1}\rangle$, $|e_{2}\rangle$) states of the individual atoms these states are represented as $|0, -1\rangle = |0, g_{1}g_{2}\rangle$,  $|2, -1\rangle = |2, g_{1}g_{2}\rangle$, $|1, 0\rangle = 1/\sqrt{2}(|1, g_{1}e_{2}\rangle + |1, e_{1}g_{2}\rangle )$, and $|0, 1\rangle = |1, e_{1}e_{2}\rangle$. From the numerical solution of the time-dependent Schrodinger equation for $\delta = \pm\sqrt{6}g_{0}$ we find that all the coefficients except for the four states mentioned in Eq. (\ref{twoatomstate}) remain zero or negligible thus validating the use of only four states in Eq. (\ref{twoatomstate}). Therefore, by considering only the four above mentioned states in the dynamics of two-atom system,  we find from our analysis that four coefficients are given by (see Appendix B for derivation)
\begin{eqnarray}
a_{0}(t) & = & \cos\left (\sqrt{\frac{2}{3}}qt\right) \nonumber \\
a_{2}(t) & = & \sqrt{\frac{1}{3}}\sin\left (\sqrt{\frac{2}{3}}qt\right) \nonumber \\
b_{1}(t) & = & \sqrt{\frac{1}{2}}\sin\left (\sqrt{\frac{2}{3}}qt\right) \nonumber \\
d_{0}(t) & = & \sqrt{\frac{1}{6}}\sin\left (\sqrt{\frac{2}{3}}qt\right) 
\label{twoatomcoeff}
\end{eqnarray}
with $q = \epsilon \left (2 + \delta \right )/8$. It should be noted here that the non-zero values of the coefficients $a_{0}(t)$ and $d_{0}(t)$ result in generation of coherent superposition of atomic states $|g_{1}g_{2}\rangle$ and $|e_{1}e_{2}\rangle$ at time $t > 0$. To see this more explicitly we write the atomic density matrix from the combined density matrix $\rho_{com}(t) = |\Psi_{com}(t)\rangle\langle\Psi_{com}|$ by tracing over the field state as
\begin{equation}
\rho_{atom}(t)  =  \left (|a_{0}(t)|^{2} + |a_{2}(t)|^{2}\right )\rho_{-1-1} + |b_{1}(t)|^{2}\rho_{00} + |d_{0}(t)|^{2}\rho_{11} + a_{0}(t)d_{0}^{*}(t)e^{2it}\rho_{-11} + a_{0}^{*}(t)d_{0}(t)e^{-2it}\rho_{1-1}
\end{equation}
with $\rho_{PQ} = |P\rangle\langle Q|$ and $P, Q = 1, 0,-1$ denoting atomic density matrices. The last two non-diagonal terms of the atomic density matrix $\rho_{atom}(t)$ arise due to coherent superposition of atomic states $|g_{1}g_{2}\rangle$ and $|e_{1}e_{2}\rangle$ and these terms signify the presence of the two-particle correlation in the atomic system. It is well known that presence of such two-particle correlation is essential for the generation of atomic squeezed states \cite{arup1,zhou,ficek} and this also makes atomic squeezing parameter a reliable measure of entanglement \cite{cirac} for this kind of states. This clearly shows that in the atom-photon interaction the two-photon correlation of squeezed state is transferred to the collection of atoms thereby producing correlated atomic states. To verify the generation of atomic squeezed state we calculate all the required averages of various components of angular momentum in the state given by Eq. (\ref{twoatomstate}). First, we find that direction of   $\langle\vec{J}\rangle$ does not change with time and it remains aligned along initial z-direction and we obtain following expressions for $\langle J_{z}\rangle$ and $\langle J_{y}^{2}\rangle$. 
\begin{equation}
\langle J_{z}(t)\rangle = -\left ( 1 - \frac{5}{6}\sin^{2}\left (\sqrt{\frac{2}{3}}qt\right )  \right )
\end{equation}

\begin{equation}
\langle J_{y}^{2}(t)\rangle = \frac{1}{2}\left ( 1  + \frac{1}{2}\sin^{2}\left (\sqrt{\frac{2}{3}}qt\right ) - \frac{2}{\sqrt{6}}\sin\left (\sqrt{\frac{2}{3}}qt\right ) \cos\left (\sqrt{\frac{2}{3}}qt\right )\cos 2t  \right )
\end{equation}

\begin{figure}[h]
\hspace{-0.0cm}
\includegraphics [scale=0.7]{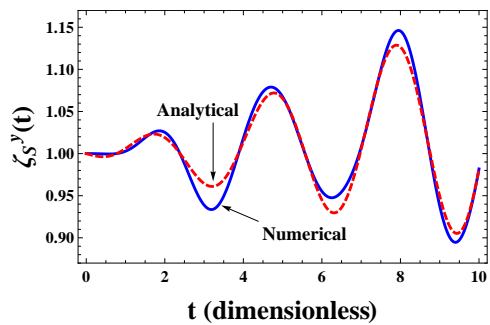}
\caption{(color online) Plot of analytical (dashed line) and numerical (solid line) solution of the squeezing parameter $\zeta_{S}^{y}(t)$ versus dimensionless time ($t$) for $\epsilon=0.05$, $\Omega=2+\sqrt{6}g_{0}$ and $g_{0}=0.5$. Here, we assume the atoms ($N=2$) to be initially in their ground states and the cavity field mode in the vacuum state.}
\end{figure}\label{fig2}

In accordance with the definition of atomic squeezing parameter we calculate the variance of the y-component of the angular momentum which is normal to the mean spin vector $\langle\vec{J}\rangle$ and the corresponding squeezing parameter is found to be
\begin{equation}
\zeta_{S}^{y}(t) = \left [ \frac{1 + \frac{1}{2}\sin^{2}\left (\sqrt{\frac{2}{3}}qt\right) - \frac{2}{\sqrt{6}}\sin\left (\sqrt{\frac{2}{3}}qt \right) \cos\left (\sqrt{\frac{2}{3}}qt\right) \cos 2t }{1 - \frac{5}{6}\sin^{2}\left (\sqrt{\frac{2}{3}}qt\right)}\right ]
\label{analyticalsq}
\end{equation}

To explicitly show that above expression indeed takes value less than unity we plot $\zeta_{S}^{y}(t)$ given by Eq. (\ref{analyticalsq}) as a function of $t$ in Fig. 2 along with the corresponding results obtained by numerically solving Eq.(\ref{equationstationary}) for comparison.  It can be clearly seen from Fig. 2 that $\zeta_{S}^{y}(t)$ attains value less than unity in finite interval of time clearly demonstrating the generation of atomic squeezed states. Moreover, we also observe that the numerical result for atomic squeezing  is quite close to the corresponding result obtained via analytical expression given by Eq. (\ref{analyticalsq}). Therefore, we conclude that the inclusion of only four states mentioned above for the calculation of dynamics of two-atom system in a non-stationary cavity under the condition $\Omega=2+\sqrt{6}g_{0}$ is quite accurate.  It is important to note here that non-zero value of the coefficients $a_{0}(t)$ and $d_{0}(t)$ is essential for the reduction of fluctuation $\zeta_{S}^{y}$. On the other hand, in the absence of mirror oscillation ($\epsilon = 0$) and for the same initial states of atoms and the field as considered above $\zeta_{S}^{y}(t)$ always remains unity signifying absence of spin squeezing \cite{saito}. Therefore, our analysis for the two two-level atoms clearly elucidates that how two-photon correlation from the cavity field mode is transferred to the atoms in the non-stationary cavity to generate atomic or spin squeezed states. 

Before proceeding further we wish to note here that for a two-atom system with $\delta = 0$ (as chosen for the numerical results presented in Fig. 1) it is not possible to solve the time dependent Schrodinger equation analytically due to coupling of coefficients with even number of photons leading to infinite set of differential equations. However due to generation of  states with even number of photons ($n \geq 2$) it is expected that higher spin squeezing will be generated.  

\begin{figure}[h]
\hspace{-0.0cm}
\begin{tabular}{cc}
\includegraphics [scale=0.70]{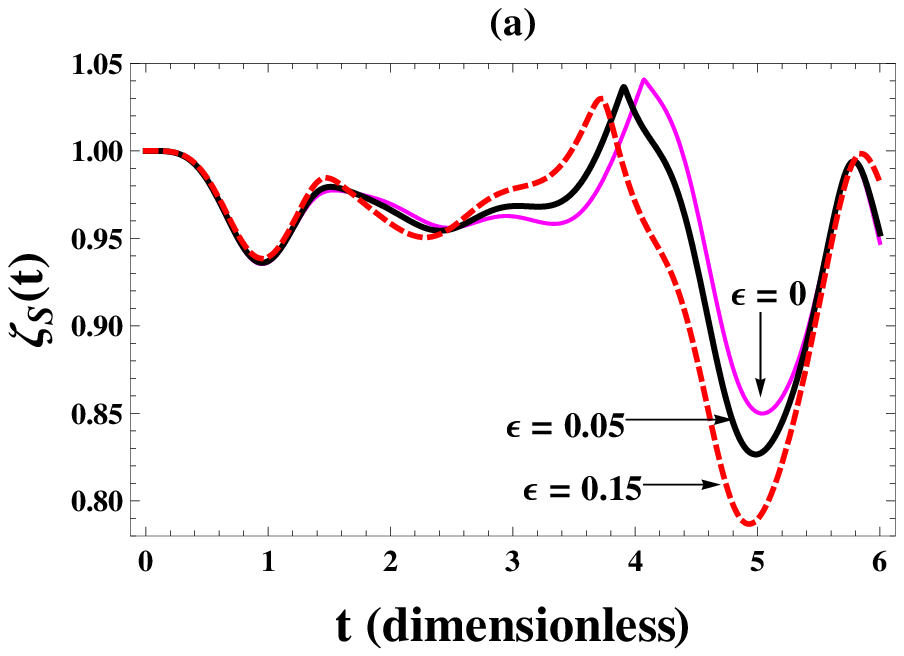}& \includegraphics [scale=0.70]{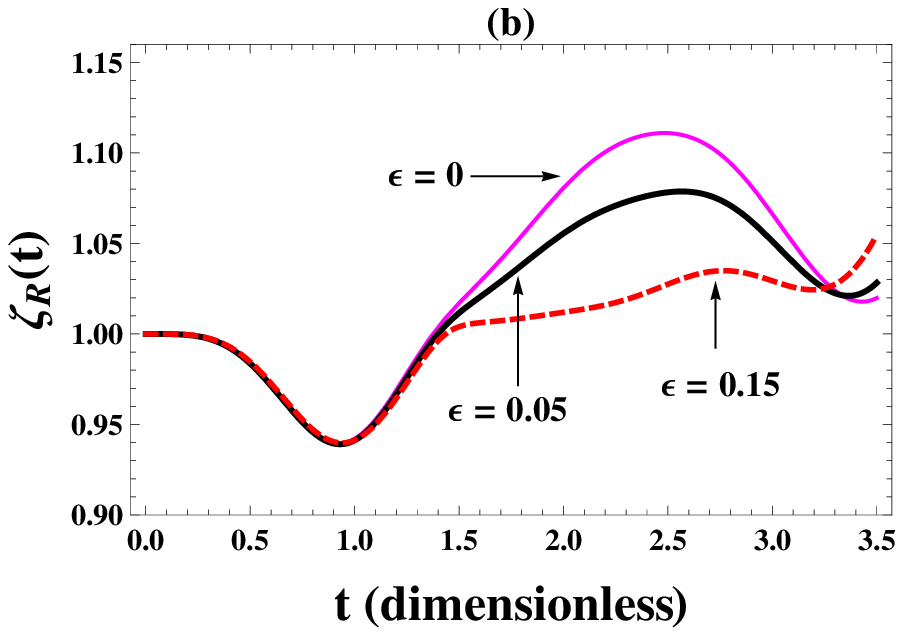}\\
\end{tabular}
\caption{(color online) (a) Plot of squeezing parameter $\zeta_{S}(t)$ as a function of dimensionless time ($t$) for three different values of modulation amplitude with $\epsilon=0$ (thin solid line), $\epsilon=0.05$ (thick solid line) and $\epsilon=0.15$ (dashed line). (b) Plot of squeezing parameter $\zeta_{R}(t)$ versus dimensionless time ($t$) for three different values of modulation amplitude with $\epsilon=0$ (thin solid line), $\epsilon=0.05$ (thick solid line) and $\epsilon=0.15$ (dashed line). Here we assume $\psi(0)=|J,-J>$ ($J=10$) and the cavity field to be initially prepared in a coherent state with an average of one quanta. The other parameters used are $g_{0}=0.6$, $\Omega=2$ and $\omega_{a}=1$.}
\end{figure}\label{fig3}

Having discussed the generation of atomic squeezed state in a non-stationary cavity with the field mode initially in the vacuum state we now consider the case when the cavity mode is initially prepared in a coherent state. In this connection we mention here that atomic squeezing property of a system of N two-level atoms interacting with a single radiation mode of a stationary cavity (Tavis-Cummings model), which is initially in a coherent state has already been reported in Ref. \cite{saito}. Thus it would be interesting to investigate the effect of cavity oscillations, which gives rise to extra non-linear terms in the Hamiltonian, on the squeezing property of atoms with the cavity mode in a coherent state. For this purpose once again we solve coupled differential equation of Eq. (\ref{eq8}) numerically by using Mathematica 9.0.  The calculations are carried out for initial atomic state $|\psi(0)\rangle = |J,-J>$ with $J = 10$ and for the field mode in a coherent state $|\alpha\rangle$ ($c_{n} = e^{-|\alpha|^{2}/2}\alpha^{n}/\sqrt{n!}$) with $|\alpha| = 1.0$. The results of these calculations for $\zeta_{S}(t)$ and $\zeta_{R}(t)$ are shown in Figs. 3(a) and 3(b) respectively, where we plot squeezing parameters as functions of dimensionless time ($t$) for three different values of modulation amplitude $\epsilon$: $\epsilon=0$ (thin solid line), $\epsilon=0.05$ (thick solid line) and $\epsilon=0.15$ (dashed line). Note that the result for $\epsilon = 0$ corresponds to the stationary cavity case described by Tavis-Cummings model. First, we observe that for all the values of $\epsilon$ atomic squeezed states are generated in the time range $0 < t < 1.5$ by the interaction of atoms with the cavity radiation mode irrespective of the parameter employed for characterizing atomic squeezing.  The comparison of results for different values of $\epsilon$ clearly reveals that within this time range atomic squeezing characterized by $\zeta_{S}(t)$ and  $\zeta_{R}(t)$ show very little variation with respect to change in $\epsilon$. On the other hand, for longer time  scales ($t > 1.5$) two parameters exhibit distinctly different atomic squeezing behavior. For example, in the longer time scale $\zeta_{S}(t)$ attains value less than unity signifying generation of atomic squeezing and degree of squeezing increases with higher values of $\epsilon$. In contrast to this in long time scale $\zeta_{R}(t)$ remains more than unity  indicating absence of atomic squeezing. 

Furthermore, we note here that comparison of results presented in Fig. 1 with the corresponding results in Fig. 3 also shows that for field mode initially prepared in vacuum state atomic squeezing is achieved for significantly longer time scale as compared to the case when field mode is initially prepared in a coherent state. Consequently, for the vacuum case unlike normal cavity QED experiment with cavity field in a coherent state the stringent requirement of the precise control of interaction time for the generation of atomic squeezed state is not necessary \cite{saito}. 

The results discussed above on the generation of atomic squeezed state by DCE can be realized with the help of practical schemes recently proposed for the generation and detection of DCE photons \cite{yamamoto,onofrio}. As mentioned before both the schemes have proposed to use the cavities operating in the GHz regime. Therefore, the presence of thermal photon will affect the results discussed above. In order to reduce the effect of thermal photon on the generation of squeezed photons the cavity should be cooled below 100 mK. In a realistic cavity-QED system, decay of the cavity mode will also degrade the degree of field squeezing and this in turn affect the magnitude of atomic squeezing. However, the loss of photons through the cavity mirrors can be minimized by using a high-Q cavity. For example, it has been shown in Ref. \cite{yamamoto} that using a semiconductor plasma mirror \cite{naylor} with Q of the order $10^{3}$ it is possible to achieve the so-called threshold condition for generation of squeezed photons. Therefore, we expect that a beam of Rydberg atoms interacting with the field mode of this cavity will be able to produce significant amount of atomic squeezing. Finally, the atomic squeezed states generated in the non-stationary cavity can be experimentally detected  by performing  measurement of atomic state population using Ramsey separated field method on the Rydberg atoms \cite{wine1}.   

\section{Conclusion}

In this paper we have studied the effect of periodic time modulation of cavity frequency on the generation of atomic or spin squeezed states. For this purpose we consider a system of N two-level atoms interacting with a single radiation mode of a non-stationary cavity with harmonically vibrating wall. We demonstrate that by modulating the cavity mode it is possible to generate atomic squeezed state by allowing an ensemble of atoms prepared in their ground states to interact with the cavity mode initially in vacuum state. In the absence of modulation, no atomic squeezing is observed. Like squeezed photon generation by DCE the efficient generation of atomic squeezed states occurs when modulation frequency of the cavity mode is twice its own frequency.  By studying a simpler system of two two-level atoms confined in a non-stationary cavity we explicitly show that atomic squeezing arises due to generation of coherent superposition of atomic states  by transfer of two-photon correlation from field mode to the atoms. For this two-atom system we also derive analytical expression for atomic squeezing for a particular value of detuning between the modulation frequency and the cavity mode by considering dynamics of few essential states. We also study the effect of modulation on the generation of atomic squeezed state for the case in which the cavity mode is initially prepared in a coherent state. Our study clearly reveals that for the cavity mode initially prepared in vacuum state the degree of atomic squeezing  increases with the increase in modulation amplitude. Therefore, modulation amplitude acts as an additional handle for controlling the squeezing of spins.

\section{Acknowledgements}

Neha Aggarwal and A. Bhattacherjee acknowledge financial support from the Department of Science and Technology, New Delhi for financial assistance vide grant SR/S2/LOP-0034/2010. Arup Banerjee wishes to thank Dr. P. K. Gupta for his encouragement and support and Mr. Krishnkanta Mondal for his help.

\section{Appendix A}

In general, a single-mode quantized cavity field is equivalent to a harmonic oscillator of unit mass such that its time-dependent Hamiltonian becomes:

\begin{equation}\label{c1}
H_{c}=\frac{1}{2}(\omega_{c}^{2}(t) Q^2+P^2),
\end{equation}

where, $Q$ and $P$ are hermitian operators and play the role of canonical position and momentum respectively such that $[Q,P]=i$. Here, $\omega_{c}(t)$ represents the time-dependent cavity frequency. The operators $Q$ and $P$ can be written in terms of creation ($a^{\dagger}$) and annihilation ($a$) operators as:

\begin{equation}
Q=\frac{1}{\sqrt{2\omega_{c}(t)}}(a+a^{\dagger}),
\end{equation} 

\begin{equation}
P=i\sqrt{\frac{\omega_{c}(t)}{2}}(a^{\dagger}-a),
\end{equation}

where the operators $a$ and $a^{\dagger}$ satisfy the commutation relation $[a,a^{\dagger}]=1$. As a result, the equations of motion for $a$ and $a^{\dagger}$ take the following form:

\begin{equation}\label{c2}
\dot{a}=-i \omega_{c}(t)a+\frac{1}{2 \omega_{c}(t)}\frac{d \omega_{c}(t)}{dt}a^{\dagger},
\end{equation}

\begin{equation}\label{c3}
\dot{a}^{\dagger}=i \omega_{c}(t)a^{\dagger}+\frac{1}{2\omega_{c}(t)}\frac{d \omega_{c}(t)}{dt}a.
\end{equation}

Thus, both the equations for $a$ and $a^{\dagger}$ are coupled to each other. Therefore, the form of Hamiltonian in terms of $a$ and $a^{\dagger}$ correctly generating the above equations of motion (\ref{c2}) and (\ref{c3}) can be written as:

\begin{equation}
H_{c}=\omega_{c}(t)a^{\dagger}a+i\xi(t)(a^{\dagger 2}-a^{2}),
\end{equation}  

where $\xi(t)=\frac{1}{4\omega_{c}(t)}\frac{d \omega_{c}(t)}{dt}$. This last term in the above Hamiltonian is responsible for parametric amplification which helps in generating the squeezed states of the optical cavity field \citep{scully}.

\section{Appendix B}
In this Appendix we present derivation of Eq. (\ref{twoatomcoeff}). For this  purpose we first go to the interaction picture by transforming $|\Psi_{com}(t)]\rangle = exp[-it\Omega/2(a^{\dagger}a + J_{z})]|\Phi(t)\rangle$. Under this transformation and by using rotating wave approximation, we get the following expression for the transformed Hamiltonian from Eq. (\ref{eq1}):
\begin{equation}\label{eqapp1}
H = \left (\Delta - \delta/2\right )J_{z} - \delta/2 a^{\dagger}a + g_{0}\left (aJ_{+} + a^{\dagger}J_{-}\right ) + iq(a^{\dagger 2}-a^{2}).
\end{equation}
where $\Delta = \omega_{a} - \omega_{0}$, $\delta = \Omega/2 - \omega_{0}$, and $q =  \epsilon \left (2 + \delta \right )/8$. For further analysis we choose $\omega_{0} = 1$. The above Hamiltonian has been obtained for $\epsilon << 1$. We now use the above Hamiltonian for a two-atom system whose state  space is spanned by the states $|J, M\rangle$ with $J = 1$ and $M = -1, 0, 1$. For this system we expand the wave function $|\Phi(t)\rangle$ as (for $\Delta = 0$)
\begin{equation}
|\Phi(t)\rangle = \sum_{n = 0}^{\infty}e^{int\delta/2}\left (a_{n}(t)e^{-it\delta/2}|n,-1\rangle + b_{n}(t)|n,0\rangle + d_{n}(t)e^{it\delta/2}|n,1\rangle\right )
\label{wavefunction}
\end{equation}
where $|n, M\rangle$ denotes the combined state of $n$ photons and the two-atom system in collective state $|M\rangle$. For notational convenience we denote collective atomic state $|J, M\rangle$ by just $|M\rangle$. By substituting $|\Phi(t)\rangle$ given by Eq. (\ref{wavefunction}) in time-dependent Schrodinger equation we arrive at following equations for the coefficients:
\begin{eqnarray}
\dot{a}_{n}(t) & = & -ig_{0}\sqrt{2n}b_{n-1}(t) + q\left (\sqrt{n(n - 1)}e^{-i\delta t}a_{n - 2}(t) - \sqrt{(n + 1)(n + 2)}a_{n + 2}(t)e^{i\delta t}\right ) \nonumber \\
\dot{b}_{n-1}(t) & = & -ig_{0}\sqrt{2n}a_{n}(t) - ig_{0}\sqrt{2(n - 1)}d_{n-2}(t) + q\left (\sqrt{(n - 1)(n - 2)}e^{-i\delta t}b_{n - 3}(t) - \sqrt{n(n + 1)}b_{n + 1}(t)e^{i\delta t}\right ) \nonumber \\
\dot{d}_{n-2}(t) & = & -ig_{0}\sqrt{2(n - 1)}b_{n-1}(t) + q\left (\sqrt{(n - 2)(n - 3)}e^{-i\delta t}d_{n - 4}(t) - \sqrt{n(n + 1)}d_{n}(t)e^{i\delta t}\right ) 
\label{equationstationary}
\end{eqnarray}
In order to solve these coupled differential equations we adopt the method used in Refs. \cite{dodonov20121,dodonov20122,dodonov20123} in the weak modulation limit defined by $\epsilon << g$. In accordance with this method we first solve the above equations for $q = 0$ (stationary cavity case) and get
\begin{eqnarray}
b_{n-1}(t) & = & A_{n}e^{-i\Omega_{n}t} + B_{n}e^{i\Omega_{n}t}  \nonumber \\
a_{n}(t) & = & \frac{g_{0}\sqrt{2n}}{\Omega_{n}}\left ( A_{n}e^{-i\Omega_{n}t} - B_{n}e^{i\Omega_{n}t} - C_{n} \right ) \nonumber \\
d_{n-2}(t) & = & \frac{g_{0}\sqrt{2(n - 1)}}{\Omega_{n}}\left ( A_{n}e^{-i\Omega_{n}t} - B_{n}e^{i\Omega_{n}t} + \frac{n}{n - 1}C_{n} \right ) 
\label{soutionstationary}
\end{eqnarray}
where $\Omega_{n} = g_{0}\sqrt{2(2n - 1)}$ and the constants $A_{n}$,$B_{n}$, $C_{n}$  are determined by initial conditions. For $q\neq 0$, the solutions given by Eq. (\ref{soutionstationary}) are once again substituted into Eq. (\ref{equationstationary}) assuming that the coefficients $A_{n}$,$B_{n}$, and $C_{n}$ are slowly varying function of time to obtain differential equations for these coefficients.  It is easy to check that for specific values of detuning $\delta$ some of these coefficients are multiplied by exponential factors with arguments larger than $q$ and these terms are neglected by invoking rotating wave approximation. On the other hand, coefficients, which are multiplied by time independent factors are retained to obtain simplified effective dynamics. We find that for initial state $|\Psi_{com}(0)\rangle = |0, -1\rangle = |0,g_{1}g_{2}\rangle$ and $\delta = g_{0}\sqrt{6}$ at most two photons can be created. Under this condition the coefficient $a_{0}(t)$ associated with the initial state mentioned above satisfy following coupled differential equations 
\begin{eqnarray}
\dot{a}_{0}(t) & = & -q\sqrt{\frac{4}{3}}A_{2}(t) \nonumber \\
\dot{A}_{2}(t) & = &  q\sqrt{\frac{1}{3}}a_{0}(t)
\label{analytical1}
\end{eqnarray}
and the non-zero coefficients (terms of the order of $(\epsilon/g_{0})^{2}$ are neglected) are obtained via following relations:
\begin{eqnarray}
a_{2}(t) & = & \sqrt{\frac{2}{3}}A_{2}(t) \nonumber \\
b_{1}(t) & = & A_{2}(t)  \nonumber \\
d_{0}(t) & = & \sqrt{\frac{1}{3}}A_{2}(t)
\label{analytical2}
\end{eqnarray}
Using Eqs. (\ref{analytical1}) and (\ref{analytical2}) we obtain the solution given by Eq. (\ref{twoatomcoeff}).


\begin{thebibliography}{plain}
\bibitem{cklaw}C. K. Law, Phys. Rev. A \textbf{49}, 433 (1994).
\bibitem{dodonov1995} V. V. Dodonov, Phys. Lett. A \textbf{207}, 126 (1995).
\bibitem{dodonov1996} V. V. Dodonov, and A. B. Klimov, Phys. Rev. A \textbf{53}, 2664 (1996)
\bibitem{plunien2000}G. Plunien, R. Schutzhold, and G. Soff, Phys. Rev. Lett. \textbf{84}, 1882 (2000).
\bibitem{crocee2001}M. Crocce, D. A. R. Dalvit, and F. D. Mazzitelli, Phys. Rev. A \textbf{64}, 013808 (2001).
\bibitem{dodonov2010}  V. V. Dodonov,  Phys. Scr.  \textbf{82}, 038105 (2010)
\bibitem{nation2012}P. D. Nation, J. R. Johansson, M. P. Blencowe, and F. Nori, Rev. Mod. Phys. \textbf{84}, 1 (2012). 
\bibitem{dodonov2011} A. V. Dodonov, R. L. Nardo, R. Migliore, A. Messina, and V. V. Dodonov, J. Phys. B: At. Mol. Opt. Phys. \textbf{44}, 225502 (2011).
\bibitem{dodonov20121}A. V. Dodonov and  V. V. Dodonov, Phys. Rev. A \textbf{85}, 055805 (2012).
\bibitem{dodonov20122}A. V. Dodonov and  V. V. Dodonov, Phys. Rev. A \textbf{85}, 063804 (2012).
\bibitem{dodonov20123} A. V. Dodonov and  V. V. Dodonov, Phys. Rev. A \textbf{86}, 015801 (2012).
\bibitem{lozovik2001}A. M. Fedotov, N, B, Narozhny, and Y. E. Lozovik, Phys. Lett. A \textbf{274}, 213 (2000).
\bibitem{yamamoto} T. Kawakubo and K. Yamamoto, Phys. Rev. A \textbf{83}, 013819 (2011).
\bibitem{onofrio} W. Kim, J. H. Brownell, and R. Onofrio, Phys Rev. Lett. \textbf{96}, 20042 (2006)
\bibitem{ruby} R. Ruby, P. Bradley, J. D. Larson III, and Y. Oshmyansky, Electron. Lett. \textbf{35}, 794 (1999)
\bibitem{cleland}A. N. Cleland and M. R. Geller, Phys. Rev. Lett. \textbf{93}, 070501 (2004)
 \bibitem{kita} M. Kitagawa, M. Ueda, Phys. Rev. A \textbf{47}, 5138 (1993).
\bibitem{wine} D. J. Wineland et al., Phys. Rev. A \textbf{46}, R6797 (1992).
\bibitem{wine1}D. J. Wineland et al., Phys. Rev. A \textbf{50}, 67 (1994).
\bibitem{toth}G. Toth, C. Knapp, O.  Guehne, H. J. Briegel,  Phys. Rev. A, \textbf{79}, 042334 (2009). 
\bibitem{nori} J, Ma, X. Wang, C. P. Sun, and  F. Nori Phys. Reports \textbf{509}, 89 (2011).
\bibitem{palma} G. M. Palma and P. L. Knight, Phys. Rev. A \textbf{39}, 1962 (1989).
\bibitem{puri}G. S. Agarwal and R, R, Puri, Phys. Rev. A \textbf{50}, 67 (1994).
\bibitem{arup} A. Banerjee, Phys. Rev. A \textbf{54}, 5327 (1996).
\bibitem{vernac}L. Vernac, M. Pinard, E. Giacobino, Eur. Phys. J. D \textbf{17}, 125 (2001).
\bibitem{polzik}E. S. Polzik, Nature \textbf{453}, 45 (2008).
\bibitem{cronin}A. D. Cronin, J. Schmiedmayer, D. E. Pritchard, Rev. Modern Phys. \textbf{81}, 1051 (2009).
\bibitem{sorensen}A. Sorensen, L. M. Duan, J. I. Cirac, P. Zoller, Nature \textbf{409}, 63 (2001).
\bibitem{bigelow}N. Bigelow, Nature \textbf{409}, 27 (2001).
\bibitem{gue}O. Guehne, G. Toth, Phys. Rep. \textbf{474}, 1 (2009).
\bibitem{tur} Q. A. Turchette et al., Phys. Rev. Lett. \textbf{81}, 3631 (1998).
\bibitem{meyer}V. Meyer et al., Phys. Rev. Lett. \textbf{86}, 5870 (2001).
\bibitem{leib}D. Leibfried et al., Science \textbf{304}, 1476 (2004).
\bibitem{wang}X. Wang, B. C. Sanders, Phys. Rev. A \textbf{68}, 012101 (2003).
\bibitem{kor}J. K. Korbicz, J. I. Cirac, M. Lewenstein, Phys. Rev. Lett. \textbf{95}, 120502 (2005); J. K. Korbicz et al., Phys. Rev. A \textbf{74}, 052319 (2006).
\bibitem{yi}S. Yi, H. Pu, Phys. Rev. A \textbf{73}, 023602 (2006).
\bibitem{arecchi}F. T. Arecchi, E. Courtens, R. Gilmore, Phys. Rev. A \textbf{6}, 2211 (1972).
\bibitem{tavis} M. Tavis, F. W. Cummings, Phys. Rev. \textbf{170}, 379 (1968).
\bibitem{ulam} D. Ulam-Orgikh, and M. Kitagawa, Phys. Rev. A \textbf{64}, 052106 (2001).
\bibitem{arup1}Arup Banerjee, Preprint quant-phys/0110032, 2000.
\bibitem{zhou}L. Zhou, H. S. Song, and C. Li, J. Opt. B: Quant. Semiclass Opt. \textbf{4}, 425 (2004).
\bibitem{ficek}A. Messikh, M. R. B. Wahiddin, C. H. Pah and Z. Ficek J. Opt. B: Quant. Semiclass Opt. \textbf{6}, 289 (2004).
\bibitem{cirac}A. Sorensen, L. M. Duan, J. Cirac, and P. Zoller, Nature \textbf{409}, 63 (2001).
\bibitem{saito} H. Saito, M. Ueda, Phys. Rev. A \textbf{59}, 3959 (1999).
\bibitem{naylor}W. Naylor, S. Matsuki, T. Nishimura, and Y. Kido,  Phys. Rev. A \textbf{80}, 043835 (2009).
\bibitem{scully} M. O. Scully, M. S. Zubairy, Quantum Optics, Cambridge University Press, Cambridge (1997)





\end{thebibliography}
\end{document}